\documentstyle[11pt,newpasp]{article}

\begin{document}

\title{X-rays from Planetary Nebulae}
\author{You-Hua Chu, Mart\'{\i}n A.\ Guerrero, and Robert A.\ Gruendl}
\affil{Astronomy Department, University of Illinois, 1002 W. Green
Street, Urbana, IL 61801, USA}

\begin{abstract}

Two sources of X-ray emission are expected from planetary nebulae:
the hot central stars with $T_{eff} > 10^5$ K, and shocked
fast stellar winds at temperatures of 10$^6 - 10^7$ K.  The stellar
emission and nebular emission differ in spatial distribution and
spectral properties.  Observations of X-ray emission from PNe
may provide essential information on formation mechanisms and 
physical conditions of PNe.  X-ray emission from PNe has been 
detected by {\em Einstein} and {\em EXOSAT}, but significant
advances are made only after {\em ROSAT} became available.
The {\em ROSAT} archive contains useful observations of $\sim$80 
PNe, of which 13 are detected.  Three types of X-ray spectra are
seen.  Only three PNe are marginally resolved by the {\em ROSAT}
instruments.  In the near future, 
{\em Chandra} will provide X-ray observations with much higher 
angular and spectral resolution, and help us understand the 
central stars as well as the hot interiors of PNe.

\end{abstract}


\section{Origins of X-ray Emission from Planetary Nebulae}

Two sources of X-ray emission are expected in planetary nebulae 
(PNe).  First, the hot central stars may reach temperatures as 
high as 100,000 -- 200,000 K and emit in soft X-rays.  Second, 
in a wind-wind interaction model of PNe, the fast (1,000 -- 3,000 
km~s$^{-1}$) stellar wind can be shock-heated to 10$^6 - 10^7$ K
and emit X-rays.
These two types of X-ray emission have different spatial extent 
and spectral properties.  X-ray emission from a hot central 
star should be a point source; its spectral properties ought to 
reflect the photospheric emission expected from the star.  
X-ray emission from the shocked fast stellar wind, on the 
other hand,  should be distributed and extend toward the inner 
wall of the dense PN shell; its spectrum ought to be characterized 
by thin plasma emission, which consists of both lines and 
bremsstrahlung emission.

X-ray emission from a PN, if detected, may provide essential 
information on PN formation mechanisms.  For example, the spatial 
distribution and spectral properties of diffuse X-ray emission 
from a PN tell us the location and physical conditions of hot, 
shocked gas in the PN interior.  Diffuse X-ray emission from a PN 
interior extending toward the inner shell walls would lend strong 
support to the wind-wind interaction model.  A point source 
centered on the PN nucleus with a hard X-ray spectrum or an 
extraordinarily high X-ray luminosity may indicate a different
emission mechanism, such as an X-ray binary.

\section{X-ray Observations of Planetary Nebulae}

X-ray observations of planetary nebulae (PNe) have been carried
out by the {\em Einstein Observatory} (1978 -- 1981), the 
{\em EXOSAT} (1983 -- 1986), and the {\em ROSAT} (1990 -- 1998).
The first report of X-ray emission from a PN was made by de Korte
et al.\ (1985), using {\em EXOSAT} observations of NGC\,1360.  
Soon after, Tarafdar \& Apparao (1988) detected X-ray emission
from four PNe using archival {\em Einstein} observations, and
Apparao \& Tarafdar (1989) added another four PNe using archival
{\em EXOSAT} observations.
As these PNe are not adequately resolved by the {\em Einstein} 
and {\em EXOSAT} instruments, their X-ray emission has been 
interpreted as stellar emission.

The {\em R\"Oentgen SATellite (ROSAT)}, launched in June 1990,
carried on board three X-ray instruments with unprecedented
sensitivity and spatial resolution: two identical Position 
Sensitive Proportional Counters (PSPCs) and a High Resolution 
Imager (HRI).  The PSPCs have a $\sim$2$^\circ$ field of view, 
an on-axis angular resolution of $\sim$30$''$, and a spectral 
resolution of $\sim$45\% at 1 keV.  The HRI has a $\sim$38$'$ 
field of view, an on-axis angular resolution of $\sim$5$''$, 
but a negligible spectral resolution.  The PSPCs are sensitive 
in the energy range of 0.1 -- 2.4 keV, and the HRI 
0.1 -- 2.0 keV.

The soft X-ray response made {\em ROSAT} ideally suitable for
PN observations.  Both pointed observations and {\em ROSAT}
All-Sky Survey data were used to study X-ray emission from PNe.
Seven new detections, including three diffuse sources, were 
reported (Kreysing et al.\ 1992; Rauch, Koeppen, \& Werner 
1994; Hoare et al.\ 1995; Chu \& Ho 1995; Leahy, Kwok, \& Yin 
1998).  
Unfortunately, some of the reports of X-ray emission from PNe
were plagued by (1) misidentifications, where extraneous
background X-ray sources were identified as PN emission; (2)
erroneous interpretations of electronic ghost images in 
PSPC observations below 0.2 keV as diffuse X-ray emission; and
(3) over-interpretation of low S/N data,  of which noise
peaks were identified as diffuse emission.  Some of these
errors have been pointed out by Chu, Kwitter, \& Kaler (1993),
Hoare et al.\ (1995), Conway \& Chu (1997), and Chu, Gruendl, 
\& Conway (1998).

As the {\em ROSAT} mission ended in 1998, all {\em ROSAT} 
observations have been archived and available at the MPE in 
Germany or the HEASARC in the US.  It is now possible to 
use  the entire {\em ROSAT} archive to make a complete and 
comprehensive investigation of X-ray emission from PNe.  We
have searched for {\em ROSAT} observations that contain a
PN within the central 40$'$-diameter field of view.  We used
the list of Galactic PNe from the ``Strasbourg-ESO Catalogue of 
Galactic Planetary Nebulae" (Acker et al.\ 1992), available
from ftp://cdsarc.u-strasbg.fr/cats/V/84.  Eighty PNe have
{\em ROSAT} observations available: 17 have both PSPC and HRI
observations, 55 have only PSPC observations, and 8 have only
HRI observations.  
For each PN, we extract X-ray images from the {\em ROSAT} 
observations, and compare them to the optical image extracted 
from the Digitized Sky Survey.  A positive detection is claimed 
only if an X-ray source is centered within a PN boundary and
has no other optical counterpart, such as a foreground star or
a background AGN.  For PSPC detections, we also extract spectra 
for further analysis.  The details of our archival study of 
this complete {\em ROSAT} sample of PNe will be reported in a
paper by Guerrero et al.\ (1999).  The main results are 
summarized in the next section.

\section{X-ray Emission from Planetary Nebulae}

PNe emit only weakly in X-rays.  Of the 13 PNe detected by
{\em ROSAT} observations, all are within 2 kpc and have 
absorption column densities of N$_{\rm H} <$ 2$\times$10$^{21}$
cm$^{-2}$.  Their X-ray luminosities range approximately from 
10$^{31}$ to 10$^{33}$ erg~s$^{-1}$.

\subsection{Spectral Properties}

ROSAT PSPC observations of PNe have revealed three distinct
types of X-ray spectra (Conway \& Chu 1997).   

{\em Type 1} has the softest 
spectral energy distribution.  The detected photons are all at
energies below $\sim$0.4 keV, and the counts increase toward 
lower energies.  PNe with Type 1 spectra include NGC\,246,
NGC\,1360, NGC\,3587, NGC\,6853, K\,1-16, and A\,30.  All, 
except A\,30, are unresolved X-ray sources and have stellar 
$T_{eff} > 100,000$ K.  Their X-ray spectra can be fitted by
blackbody emission models with temperatures of $\sim150,000$ K.
It is most likely that these X-ray sources represent photospheric
emission from the central stars.

{\em Type 2} spectra are harder, with most detected photons at
energies above 0.5 keV.  PNe with Type 2 spectra include 
BD+30$^\circ$3639 and NGC\,6543, whose X-ray spectra can be fitted
by thin plasma emission models with plasma temperatures of a few
$\times$10$^6$ K.  It is interesting that both nebulae have been
reported to host diffuse X-rays.  Three additional PNe may have 
Type 2 spectra: A\,36, K\,1-27, and NGC\,7009; their PSPC spectra
are noisy but do not show the tell-tale peak toward the lowest
energy bin of the PSPC as shown in Type 1 or Type 3 spectra.

{\em Type 3} spectra are composite, with a strong soft component
and a weak hard component.  Only NGC\,7293 and LoTr\,5 belong to
this category.  Neither is resolved.  The origin of the hard 
component is difficult to explain.  LoTr\,5 is a known binary, 
but NGC\,7293 is not.  

\subsection{Diffuse X-ray Emission}

Diffuse X-ray emission has been reported in three PNe: A\,30, 
BD+30$^\circ$3639, and NGC\,6543.  In all three cases, the
X-ray emission is marginally resolved by the instruments,
with the source sizes being $\leq$1.5 times the instrumental
FWHM.  The diffuse emission surrounding the central star of
A\,30 is detected at only a 2$\sigma$ level (Chu, Chang, \&
Conway 1997).  The diffuse X-ray emission of NGC\,6543 is 
detected only by PSPC, with a $\sim$30$''$ instrumental 
resolution, which is larger than the entire ``cat's eye'' of 
NGC\,6543.  X-ray observations with a higher spatial resolution
are needed to confirm the diffuse X-ray emission in these
nebulae.  The hard X-ray components in Type 2 and Type 3 spectra
are indicative of the presence of hot gas; PNe with such spectra 
are therefore the best future targets to search for diffuse 
X-ray emission.

\section{Future Space Observations}

The {\em Chandra X-ray Observatory}, one of the four
{\em Great Observatories}, was launched on 1999 July 23 and
delivered to its working orbit on August 7.  In the very
near future, {\em Chandra} will be observing with its 
Advanced CCD Imaging Spectrometer (ACIS), High Resolution 
Camera (HRC), and transmission grating spectrometers at 
low and high energies (LETG and HETG, respectively).

In Cycle 1, four PNe will be observed with the spectroscopic
array (ACIS-S), which offers $\sim$1$''$ angular resolution 
and an energy resolution of $E/\Delta E$ = 10 at 0.5 keV.
The four PNe's names, exposure times, and PIs
are: BD+30$^\circ$3639, 20 ks, Kastner; NGC\,6543, 50 ks, Chu;
NGC\,7027, 20 ks, Kastner; and NGC\,7293, 50 ks, Chu.

The ACIS-S observations of these PNe will show unambiguously
whether the X-ray emission originates from the central stars or 
the hot, shocked gas in PN interiors.  If diffuse X-ray emission
is detected, we may use it to determine the location and 
temperature of the hot gas.  If point sources are detected,
we may perform timing analysis as well as spectral analysis
to determine whether X-ray binaries are present.

Besides {\em Chandra}, two other space observatories may
be used to study the hot star and gas in PNe: the
European {\em X-ray Multi-Mirror Mission (XMM)} and the 
{\em Far Ultraviolet Spectroscopic Explorer (FUSE)}.  
Compared to {\em Chandra}, {\em XMM} has a greater sensitivity
to soft X-rays but a worse angular resolution.  If no diffuse
X-ray emission from PNe is confirmed by {\em Chandra} or {\em XMM},
the presence of 10$^6$ -- 10$^7$ K gas in PN interiors will be
in serious doubt.  It is then important to search for cooler
gas at a few times 10$^5$ K gas.  {\em FUSE} provides a means 
to detect such gas in a PN interior via the nebular O\,VI 
absorption lines against the spectrum of the central star, if
one can distinguish among the foreground interstellar component, 
the photoionized component, and the collisionally ionized component
of O$^{+5}$.


\acknowledgments
This research is supported by NASA grant NAG 5-8103 through the
Astrophysical Data Program.

\end{document}